\documentclass[prd,showpacs, nofootinbib]{revtex4}
\usepackage{graphicx}
\usepackage{bm}
\usepackage{longtable}
\usepackage{subfigure}

\begin{document}

\title{Comment on ``On curvature coupling and quintessence fine-tuning''}

\author{Urbano Fran\c{c}a} 

\email{urbano@sissa.it}

\affiliation{SISSA / ISAS, Astrophysics Sector, 
Via Beirut 4, 34014, Trieste, Italy}

\begin{abstract}
In this comment, we show explicitly that the phenomenological model
in which the quintessence field depends linearly on the 
energy density of the spatial curvature can provide acceleration
for the universe, contrary to recent claims it could not.
\end{abstract}
                        


\maketitle


Recently Gong {\it et al.} \cite{gong05} have stated that
the phenomenological model proposed in \cite{franca05} would
not be able to provide the current acceleration of the universe. This conclusion
is based on the equation (3) of their paper, that we reproduce here,
\begin{equation} \label{eq:wrong}
\frac{\ddot{a}}{a} = - \ \frac{1}{6 m_p^2} [\rho_M + 2 \rho_R + 2 (\rho_{\phi}+ p_{\phi})] \ ,
\end{equation}
where $m_p^2 = (8 \pi G)^{-1}$, $M, R$ and $\phi$ denotes matter,
radiation and quintessence, respectively, and $a$ is the scale factor. 
As usual, $\rho_{\phi} = \dot{\phi}^2 /2 + U$, 
$p_{\phi} = \dot{\phi}^2 /2 - U$, and $U= - \rho_k \exp(\lambda \phi / m_p) 
\equiv (3 m_p^2 k/a^2) \exp(\lambda \phi / m_p)$. Dots denote
derivative with repect to cosmic time $t$. As we will show below, 
this equation is not correct, and 
therefore it leads to misleading conclusions.

The Friedmann equation is given by
\begin{equation} \label{eq:friedmann}
H^2 = \frac{1}{3m_p^2} (\rho_M +  \rho_R + \rho_k + \rho_{\phi}) \ ,
\end{equation}
and its derivative with respect to cosmic time $t$ reads
\begin{equation}
\dot{H} = \frac{\ddot{a}}{a} - H^2 = - \ \frac{1}{6m_p^2} (3 \rho_M +  4 \rho_R + 2\rho_k + 3 \dot{\phi}^2) \ ,
\end{equation}
where we had only taken the derivatives of both sides of eq. (\ref{eq:friedmann}) and 
used the fluid equations\footnote{For details about the use of 
the fluid equation for the field, see the appendix.}, $\dot{\rho}_i = - 3 H (\rho_i + p_i) $, $i = M, R, k$ and $\phi$.
Thus, we have that
\begin{equation} \label{eq:correct1}
\frac{\ddot{a}}{a} = - \ \frac{1}{6m_p^2} [\rho_M +  2 \rho_R + 2 (\dot{\phi}^2 -  U)] \ ,
\end{equation}
and since $\rho_{\phi} + 3 p_{\phi} = 2 (\dot{\phi}^2 -  U)$, this is precisely
\begin{equation} \label{eq:correct2}
\frac{\ddot{a}}{a} = - \ \frac{1}{6m_p^2} [\rho_M +  2 \rho_R + \rho_{\phi} + 3 p_{\phi}] \ ,
\end{equation}
which is the expected result, even when the spatial curvature of the universe is not zero (see 
\cite{kt90}, for instance). This result is clearly different from eq. (\ref{eq:wrong}), and
in the regime in which the field has its energy dominated by the
potential term it is able to provide the acceleration of the universe without any problems.

Therefore, since the model proposed in \cite{franca05} works, contrary to
the conclusions obtained in \cite{gong05}, there is no reason
to do the changes proposed to obtain cosmologically acceptable 
solutions that can solve the coincidence problem.

\vspace{0.5cm}

{\it Acknowledgements.} I would like to thanks 
Thomas P. Sotiriou for enlightening
discussions, and Yungui Gong for very interesting 
correspondence.
\appendix

\section{Equivalence of fluid and Klein-Gordon equations}

Conservation of the total energy-momentum tensor
requires that
\begin{equation}
\nabla_{\mu} T^{\mu \nu} = 0 \ . 
\end{equation}
For a scalar field, the energy-momentum tensor is
given by
\begin{equation}\label{eq:em1}
T_{\phi}^{\mu \nu} = \partial^{\mu} \phi \partial^{\nu} \phi 
 - g^{\mu \nu} \mathcal{L} = \partial^{\mu} \phi \partial^{\nu} \phi 
 - g^{\mu \nu} \left( \frac{1}{2} \partial_{\alpha} \phi 
\partial^{\alpha} \phi  - U\right)\ , 
\end{equation}
and in the case in which each energy-momentum tensor
is conserved separately, one can show that for $\nu =0$,
\begin{equation} \label{eq:cons}
\nabla_{\mu} T_{\phi}^{\mu \nu} = (\partial_{0}  + 
\Gamma^{\mu}_{\mu 0} ) T^{00} + \Gamma^{0}_{ii}  T^{ii} = 0\ , 
\end{equation}
where $i=1,2,3$ denotes the spatial components of the metric.
Therefore, if one uses the energy-momentum tensor for a 
perfect fluid, 
\begin{equation} \label{eq:em2}
T_{\phi}^{\mu \nu} = (\rho + p) u^{\mu}u^{\nu} - g^{\mu \nu} p \ ,
\end{equation}
equation (\ref{eq:cons}) gives the well known fluid 
equation,
\begin{equation} \label{eq:fluid}
\dot{\rho}_{\phi} + 3 H (\rho_{\phi} + p_{\phi}) = 0 \ .
\end{equation}
On the other hand, if one uses equation (\ref{eq:em1}), equation 
(\ref{eq:cons}) reads
\begin{equation} \label{eq:kg}
\ddot{\phi} + 3 H \dot{\phi} + \frac{\partial{U}}{\partial \phi} + 
\frac{\dot{a}}{\dot{\phi}}\frac{\partial{U}}{\partial a}  = 0 \ ,
\end{equation}
that is the Klein-Gordon equation of this field. Notice that it differs
from the Klein-Gordon used in \cite{gong05}, since it has the term with
the derivative with respect to the scale factor $a$. This equation
has the  ``extra'' term given in the scalar
field equation of \cite{franca05} when compared to the
scalar field equation given in \cite{gong05}.

Such a Klein-Gordon equation cannot be immediately
obtained using regular Euler-Lagrange equation, since this 
procedure requires the Lagrangian {\it to be dependent on the field $\phi$
and its first derivative $\partial_{\mu} \phi$ only}. The
{\it phenomenological} Lagragian we used in \cite{franca05} has also an
explicit dependence on $a$, through the curvature, what 
does not allow one to use regular Euler-Lagrange equation for 
deriving the equation of motion of the field. Therefore,
unless one can obtain a covariant form of
writing such a Lagrangian, the only way of obtaining the
equations of motion of the scalar field is requiring its
energy-momentum tensor to be conserved. Whether or not
such a dependence can be obtained in a realistic model is
something that was beyond the scope of \cite{franca05}, and
it is currently under investigation. However, once one
accepts it, the steps described must be followed, otherwise
one would run against inconsistencies.

Summing up, both fluid and Klein-Gordon equations have
to be consistent, equivalent, and equally able to describe 
the behaviour of the field.

\end{document}